\documentclass{aastex631}

\usepackage{comment}

\usepackage{xcolor}

\begin{document}


\title{Radio Pulsar B0950$+$08: Radiation in Magnetosphere and Sparks above Surface}

\author{Zhengli Wang}
\affiliation{Guangxi Key Laboratory for Relativistic Astrophysics, School of Physical Science and Technology, \\
Guangxi University, Nanning 530004, China}

\author{Jiguang Lu}
\affiliation{National Astronomical Observatories, Chinese Academy of Sciences, Beijing 100012, China}
\affiliation{Guizhou Radio Astronomical Observatory, Guiyang 550025, China}

\author{Jinchen Jiang}
\affiliation{National Astronomical Observatories, Chinese Academy of Sciences, Beijing 100012, China}

\author{Shunshun Cao}
\affiliation{Department of Astronomy, School of Physics, Peking University, Beijing 100871, China}

\author{Kejia Lee}
\affiliation{Department of Astronomy, School of Physics,
Peking University, Beijing 100871, China}
\affiliation{Kavli Institute for Astronomy and
Astrophysics, Peking University, Beijing 100871, China}

\author{Enwei Liang}
\affiliation{Guangxi Key Laboratory for Relativistic Astrophysics, School of Physical Science and Technology, \\
Guangxi University, Nanning 530004, China}

\author{Lunhua Shang}
\affiliation{Guizhou Normal University, Guiyang 550025, China}

\author{Weiyang Wang}
\affiliation{School of Astronomy, University of Chinese Academy of Sciences, Beijing 100049, China}
\affiliation{National Astronomical Observatories, Chinese Academy of Sciences, Jia-20 Datun Road, ChaoYang District, Beijing 100101, China}

\author{Renxin Xu}
\affiliation{State Key
Laboratory of Nuclear Physics and
Technology,
Peking University, Beijing 100871, China}
\affiliation{Department of Astronomy, School of Physics,
Peking University, Beijing 100871, China}
\affiliation{Kavli Institute for Astronomy and
Astrophysics, Peking University, Beijing 100871, China}

\author{Weiwei Zhu}
\affiliation{National Astronomical Observatories, Chinese Academy of Sciences, Beijing 100012, China}

\correspondingauthor{Jiguang Lu, Enwei Liang, Renxin Xu}
\email{lujig@nao.cas.cn, lew@gxu.edu.cn, r.x.xu@pku.edu.cn}








\begin{abstract}
We observed the nearby 100\%-duty-cycle radio pulsar B0950+08 using the Five-hundred-meter Aperture Spherical radio Telescope (FAST).
We obtained the polarization profile for its entire rotation, which enabled us to investigate its magnetospheric radiation geometry and the polar cap sparking pattern.
After we excluded part of the profile in which the linear polarization factor is low ($\lesssim 30$\%) and potentially contaminated by position angle jumps, the rest of the polarization position angle swing fits a classical rotating vector model (RVM) well.
The bestfit RVM indicates that the inclination angle, $\alpha$, and the impact angle, $\beta$, of this pulsar, are 100.5$^{\circ}$ and $-$33.2$^{\circ}$, respectively, suggesting that the radio emission comes from two poles.
We find that, in such RVM geometry, either the annular vacuum gap or the core vacuum gap model would require that the radio emissions come from a high-altitude magnetosphere with heights from $\sim 0.25~R_{\rm LC}$ to $\sim 0.56~R_{\rm LC}$, with $R_{\rm LC}$ being the light cylinder radius.
Both the main and inter-pulses' sparking points are located away from the magnetic pole, which could be relevant to the physical conditions on the pulsar surface.
\end{abstract}
\keywords{Individual pulsar --- PSR B0950$+$08 --- Magnetospheric geometry --- Radio emission -- Emission zone} 

\section{Introduction} \label{sec0}
 
%
The nearby $\sim 253\,$ms pulsar, PSR B0950$+$08 has been observed by several groups \citep[e.g.,][]{2022A&A...658A.143B,2012AJ....144..155S,2005MNRAS.364.1397J, 2004MNRAS.353.1311H,1981ApJ...249..241H}.
%
%
%
\citet{1981ApJ...249..241H} pointed out that the separation between the interpulse and main pulse is frequency-independent below 5\,GHz and the ``low level emission'' (the emission from the ``bridge'' component of the interpulse to the main pulse) of this pulsar is detected over at least $83\%$ of the rotation period, and they argued that these radio emission features could be the result of the magnetic multipole field.
%
Recently, \citet{2022MNRAS.517.5560W} found that the radio signal of this pulsar is detected over its entire rotation using FAST observations.
\citet{2012AJ....144..155S} presented the detection of its giant pulse emission and showed that the cumulative intensity density distribution of the detectable giant pulses is a power law function with index $-$2.2.
\citet{2022A&A...658A.143B} argued that the $\mu$s scale fluence variability of single pulse of this pulsar at low frequency may be caused by the diffractive scintillation.
\citet{2001ApJ...553..341E} proposed an orthogonal rotator for PSR B0950$+$08 and explained the radiation region of the main and inter pulses as the opposite magnetic poles.
%
%
While the emission mechanism and its location within a pulsar are open questions, polarization studies can help gain insights into the same.

More than half a century since pulsar discovery, there is yet no unified understanding of particle creation and acceleration in the magnetosphere ~\citep{2018PhyU...61..353B}. 
The challenges may be the poorly understood physics of pulsar surface, the vacuum gap model requires strong binding yet the space-charge-limited flow model needs weak binding on the pulsar surface.
The inner vacuum accelerators are proposed and developed by \citet[][hereafter RS75]{1975ApJ...196...51R} at first, with polar cap radius on pulsar surface, $r_p = R\sqrt{R\Omega/c}\simeq (10^4~{\rm cm})~R_6^{3/2}P^{-1/2}$, for an aligned rotator of radius $R=(10^6~{\rm cm})~R_6$ and angular frequency $\Omega=2\pi/P$ (the light speed, $c$).
Motivated by explaining both the radio and the $\gamma$-ray pulse profiles of rotation-powered pulsars and by understanding naturally the global current flows in pulsar magnetospheres, \citet{2004ApJ...606L..49Q} proposed an annular gap (AG) model.
The open-field-line magnetosphere would then be divided into two regions: the magnetic field lines in an AG region intersect the null charge surface (NCS), while the rest is called a core gap (CG).
%
%
Certainly, it is not well understood which region dominates the radiation, and a precise analysis of the emission zones of this bright pulsar should provide insight into the pulsar radiation mechanism.

Besides, the emission physics of radio pulsars is closely relevant to the polarization emission properties. Firstly, unraveling the magnetospheric geometries of the pulsars depends on the inclination angle between the magnetic and rotation axes. Secondly, the radio telescope can detect the radiation behaviors of the pulsars such as the average pulse profile, depending on the viewing angle. Finally, the detection of the radiation in the magnetosphere and the mapping of the polar cap sparking also require the study of the polarization behaviors. Therefore, a detailed study of the polarization behaviors provides global knowledge for understanding the radiation mechanism.

We describe the 110-min polarization-calibrated data observed with FAST and the data reduction in section \ref{sec1}.
To obtain the polarization emission properties of PSR~B0950$+$08 with radiation over the whole $360^{\circ}$ of longitude, we propose two tentative methods of baseline intensity determination in section \ref{sec2}.
The polarization position angle (PPA) is fitted in the classical rotating vector model~\citep[RVM,][]{1969ApL.....3..225R}, and
the related results are presented in section \ref{sec3}. Discussions and conclusions are summarized in sections \ref{sec5} and \ref{sec6}, respectively.

\section{observation and data reduction} \label{sec1}

In this work, PSR B0950$+$08 was observed with tracking mode on MJD 59820 (2022 August 29) using the 19-beam receiver system of FAST \citep{2019SCPMA..6259502J,2020RAA....20...64J}. The raw data were recorded in the 8-bit-sampled search mode \textsc{PSRFITS} format \citep{2004PASA...21..302H} with 4096 frequency channels, and the frequency resolution is $ \sim $0.122\,MHz. The entire polarization-calibrated observation is 110 minutes and the time resolution of the recorded data is $49.152\,\mu $s. To obtain the emission properties of single pulses of this pulsar, the \textsc{DSPSR} software package was adopted in the data reduction process \citep{2011PASA...28....1V}. The option ``-s'' provided by \textsc{DSPSR} was used to fold the raw data with a resampling time of $ \sim $0.25\,ms. In order to obtain accurate polarization calibration solutions, the polarization calibration signals with 30\,s were injected after each sub-integration for 30 minutes. Besides, the radio frequency interference (RFI) is eliminated using the frequency-time dynamics spectrum.


\section{baseline intensity determination} \label{sec2}

Conventional baseline subtraction is used to subtract the weak emission region away from the strong emission region (i.e., the main pulse ) of the pulsar, but it would not be suitable for PSR B0950$+$08 since the radio signal of this pulsar is detected over the whole pulse phase \citep{2022MNRAS.517.5560W}. To detect the baseline intensity of this pulsar, two tentative baseline intensity determinations are proposed.

\begin{figure}
    \centering
    \includegraphics[width = .8\linewidth]{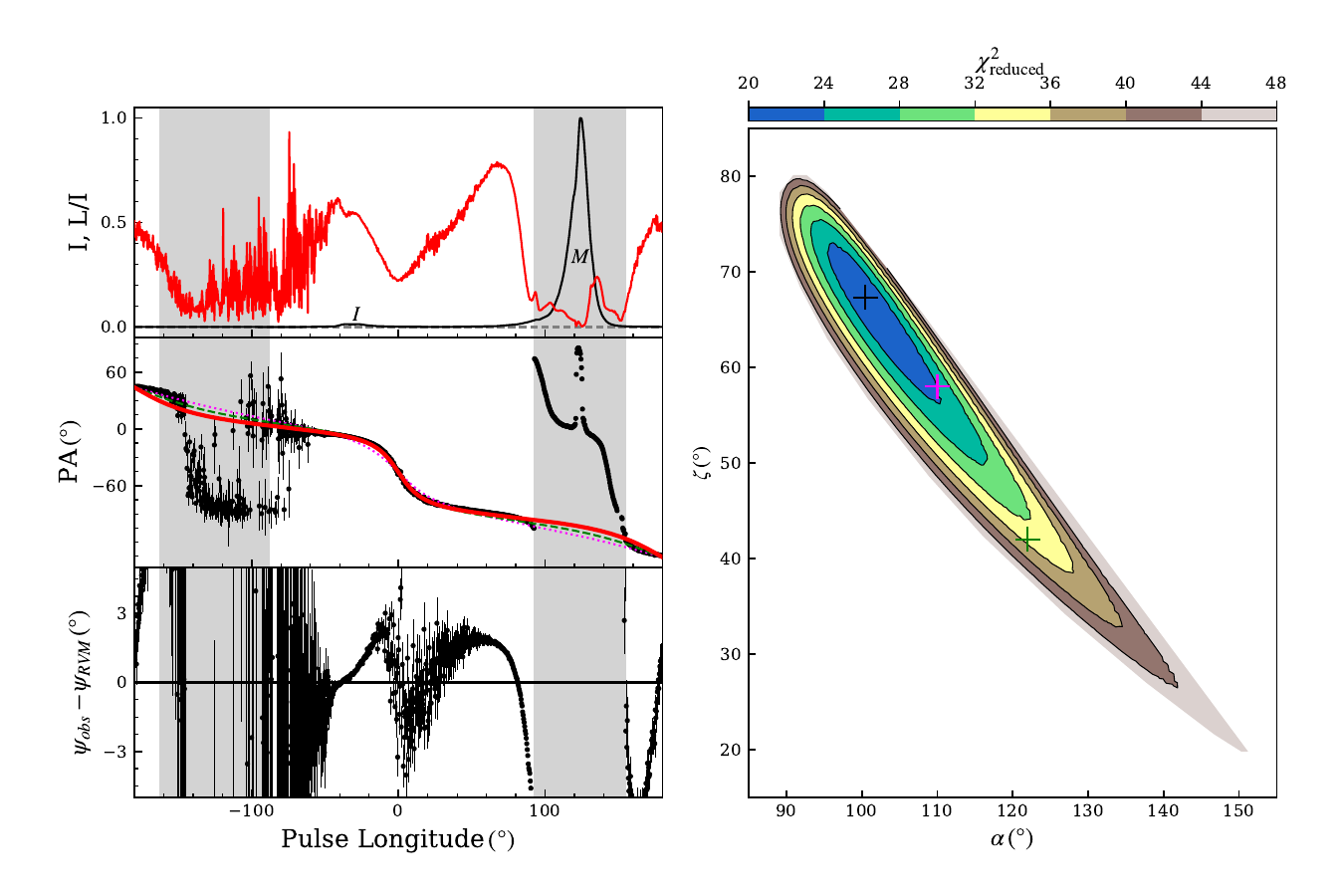}
    \caption{The RVM fit for the pulse longitudes whose linear fractions are higher than $ \sim 30 \% $ of the whole pulse phase radiative pulsar, PSR B0950$ + $08. The left-hand panels describe the total emission intensity (black curve) and the linear fraction (red curve). The observed PPAs and best RVM fit (red line) are shown in the middle panel, and it gives the inclination angle $ \alpha $ and the impact angle $ \beta \,(\beta = \zeta - \alpha)$ are $ 100.5^{\circ} $ and $ -33.2^{\circ} $, respectively. The errors between the RVM fit curve and observed PPAs are plotted in the bottom panel and whose values fall in the range from $-5^{\circ}$ to $5^{\circ}$ are believed to be the RVM intervals. Two vertical grey regions are unweighted in the fit because the left-hand vertical grey region indicates the bridge component whose polarization emission property is detected with large uncertainty. The right-hand grey region is the main pulse region whose polarization emission displays depolarization and position angle jump behaviors, and these complex polarization emission phenomena fail to be described by the RVM. The $ I $ and $ M $ indicate the locations of the peak of the interpulse and main pulse in the top panel, respectively. The steepest gradient of the RVM fit curve has been centered. The right-hand panel shows the $ \alpha - \zeta $ plane, which depicts the value of $ \chi_{\mathrm{reduced}}^2 $ from the fitting routine, and the cross denotes the location of the minimum of the $ \chi_{\mathrm{reduced}}^2 $ surface. We select also the other two parameter sets of \{$\alpha,\beta$\}, two crosses (magenta and green) in the right panel and their corresponding RVM curves (magenta dotted and green dashed curves) in the left panel, indicating that the differences between the three groups of RVM solutions are minimal.}
    \label{f0}
\end{figure}

PSR B0950$+$08 is a bright pulsar located nearby, with a dispersion measure (DM) of 2.97 pc $\mathrm{cm^{-3}}$, which indicates that the average pulse profile of this pulsar is little affected by the interstellar scintillation effect. Previous observations of PSR B0950$+$08 have not detected the mode switching behavior~\citep[e.g.][]{1981ApJ...249..241H,2012AJ....144..155S,2022A&A...658A.143B,2022MNRAS.517.5560W}, so its average pulse profile is stable at a narrow observing frequency. 
In addition, the stable profile is related to the integration, and the average pulse profiles of the radio pulsars are quite stable over long integration. In this work, the emission property that the average pulse profile of this pulsar is stable over long integration is used to detect its baseline.
To obtain the stable profile of the sub-integration of this pulsar, the 110-min polarization-calibrated observation is segmented into two sub-integration with 55 minutes. Considering the relationship between the average pulse profiles of the two sub-integrations of this pulsar and its baseline intensity, we could have
\begin{equation}
    I_{e_1} - I_{b} = \kappa (I_{e_2} - I_{b}),
    \label{eq0}
\end{equation}
where $\kappa$ is a parameter that reflects the fluctuations in the radio emission intensity of this pulsar over different sub-integrations, and it would be equal to 1 for the infinite integration.
Here, $I_{e_1}$ and $I_{e_2}$ denote the average radio intensity of this pulsar over its entire $360^{\circ}$ of longitude for the first and second sub-integrations, respectively. The $I_b$ corresponds to the baseline intensity of the entire 110-min observation. A detailed derivation of Equation (\ref{eq0}) is presented in Appendix~\ref{secA}.
The result of subtracting PSR B0950$+$08's baseline intensity according to Equation (\ref{eq0}) is shown in the left-hand panels of Figure \ref{f0}.

On the other hand, a model based on the polar cap's electric field distribution properties is put forward here to help subtract the baseline. Assuming that the polar cap of a pulsar is a voltage-regular diode.
%
The spark discharge can be considered as the breakdown of the voltage-regular diode, whose breakdown voltage is assumed as $V_0$.
Consequently, the distribution of the voltage of the base and top edges of the voltage-regular diode can be regarded as $V \sim \mathcal{N}(V_0, \sigma_V^2)$. Here $V$ and $\sigma_V$ denote the voltage of the two edges of the voltage-regular diode and its standard deviation.
%
According to the property of the Gaussian white noise distribution, the fluctuation voltage distribution of the two edges of the voltage-regular caused by each spark discharge is set to $ \Delta V $, which follows $ \mathcal{N}(0, 2 \sigma_V^2)$.
Compared with the $V_0$, the $ \Delta V $ is a small variation (i.e., $\Delta V \ll V_0$). Therefore, the energy of each polar cap spark discharge can be estimated, and its distribution is $ \mathcal{N}(0, 4 V_0^2 \sigma_V^2)$.

Considering that the spark discharge is equivalent to converting the electronic field energy of a pulsar into the number of electrons $N$, the accelerating electric field (it corresponds to the component of the electric field in the pulsar magnetosphere parallel to the magnetic field) of a pulsar is hardly changed. Therefore, the energy of the accelerated electrons is thought to be a constant. Furthermore, the radiative spectrum of a group of electrons is also hardly changed. Nonetheless, its flux has a significant change since the flux depends on the number density of the electrons $n_e$. The number of electrons produced by the spark discharge can be regarded as the number density of the electrons because the electrons almost escape from the polar cap of a pulsar at the same time.
Considering that the radiation of a pulsar is a coherent radiation process, the radiation energy is proportional to $N^2$. Consequently, the radiation energy released by each spark discharge equals the square of the Gaussian distribution. In the radiation of radio pulsars, each emission phase point may include multiple spark discharge processes, so the flux distribution of this emission phase point can be estimated, which equals the sum of squares of multiple Gaussian distributions (i.e., $\chi^2$-distribution).
We assume that the distribution of the total number of the spark discharge processes follows the Poisson distribution in the polar cap of a pulsar.

Considering the properties of the $\chi^2$-distribution and the Poisson distribution, the relationship between the radio emission of this pulsar and its baseline intensity can be given that,
\begin{equation}
	(I_e - I_b) (D_e - D_b) = \kappa (S_e - S_b)^2
	\label{eq1}
\end{equation}
where $I_e$, $S_e$ and $D_e$ denote the radio emission intensity, variance, and central moment of the third order of this pulsar, respectively. The intensity, variance, and central moment of the third order of the baseline of this pulsar are denoted as $I_b$, $S_b$, and $D_b$, respectively. Here $\kappa$ corresponds to a parameter that reflects the fluctuations in the radio emission intensity of this pulsar over different sub-integrations. The result is consistent with Equation (\ref{eq0}).

\begin{figure}[ht!]
    \centering
    \includegraphics[width = .8\linewidth]{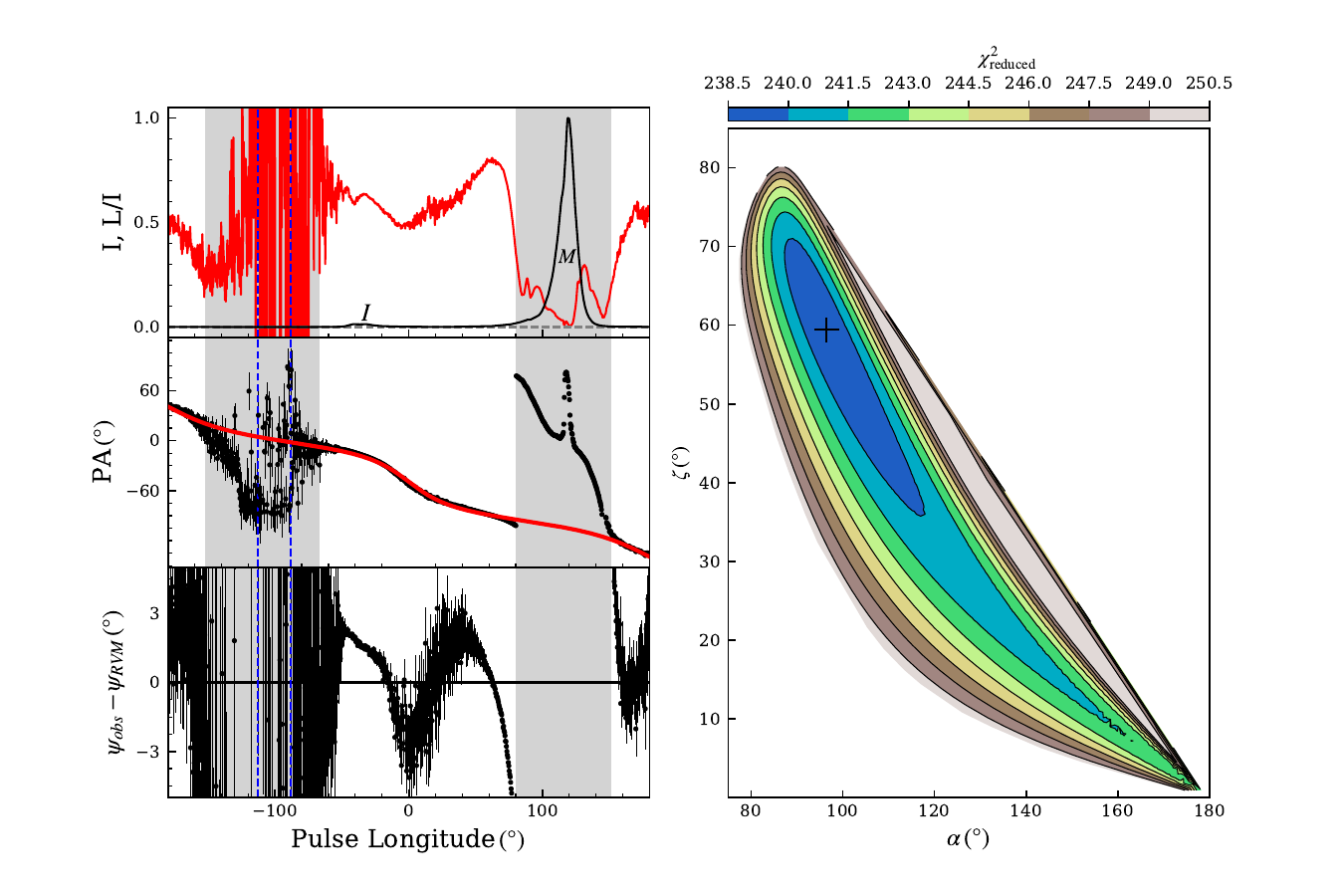}
    \caption{Same as Figure \ref{f0}, but the region between the two vertical dashed blue lines is regarded as the reference of the baseline. From the right-hand panel, it indicates that the inclination angle $\alpha = 96.5^{\circ}$ and the impact angle $\beta = -37.1^{\circ} \, (\beta = \zeta - \alpha)$. It should be noticed that the values of $ \chi_{\mathrm{reduced}}^2 $ are about ten times as much as the values of the $ \chi_{\mathrm{reduced}}^2 $ given in the right-hand panel of Figure \ref{f0}.} 
    \label{f1}
\end{figure}

Meanwhile, the conventional baseline subtraction is also taken into account. Considering that the radio emission property of PSR B0950$+$08 \citep[][]{2022MNRAS.517.5560W}, the bridge component between the main pulse and interpulse accounts for $\sim 5 \%$ of the rotating period of this pulsar is regarded as its baseline intensity. The result is shown in Figure \ref{f1}. The legends and indications are the same as Figure \ref{f0}, but for taking the bridge component between the two vertical blue lines as the reference of the baseline. In comparison to the left-hand panel of Figure \ref{f0}, there are noticeable differences in the PA-swings of the weak emission regions, such as the precursor component of the main pulse and the postcursor component of the interpulse. Further discussions on these differences will be presented in later sections.

\begin{figure*}[ht] 
\begin{minipage}[t]{0.4\linewidth}
  \centerline{\includegraphics[width=1.0\textwidth]{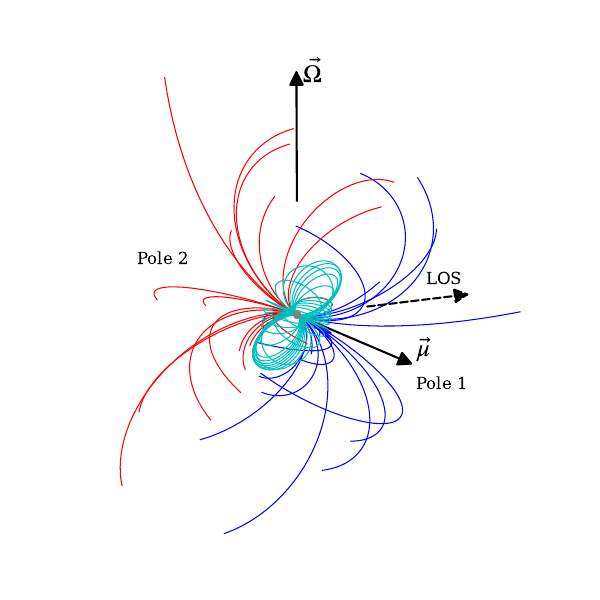}}
\centerline{(a)}
\end{minipage}
\begin{minipage}[t]{0.6\linewidth}
\centerline{\includegraphics[width=1.0\textwidth]{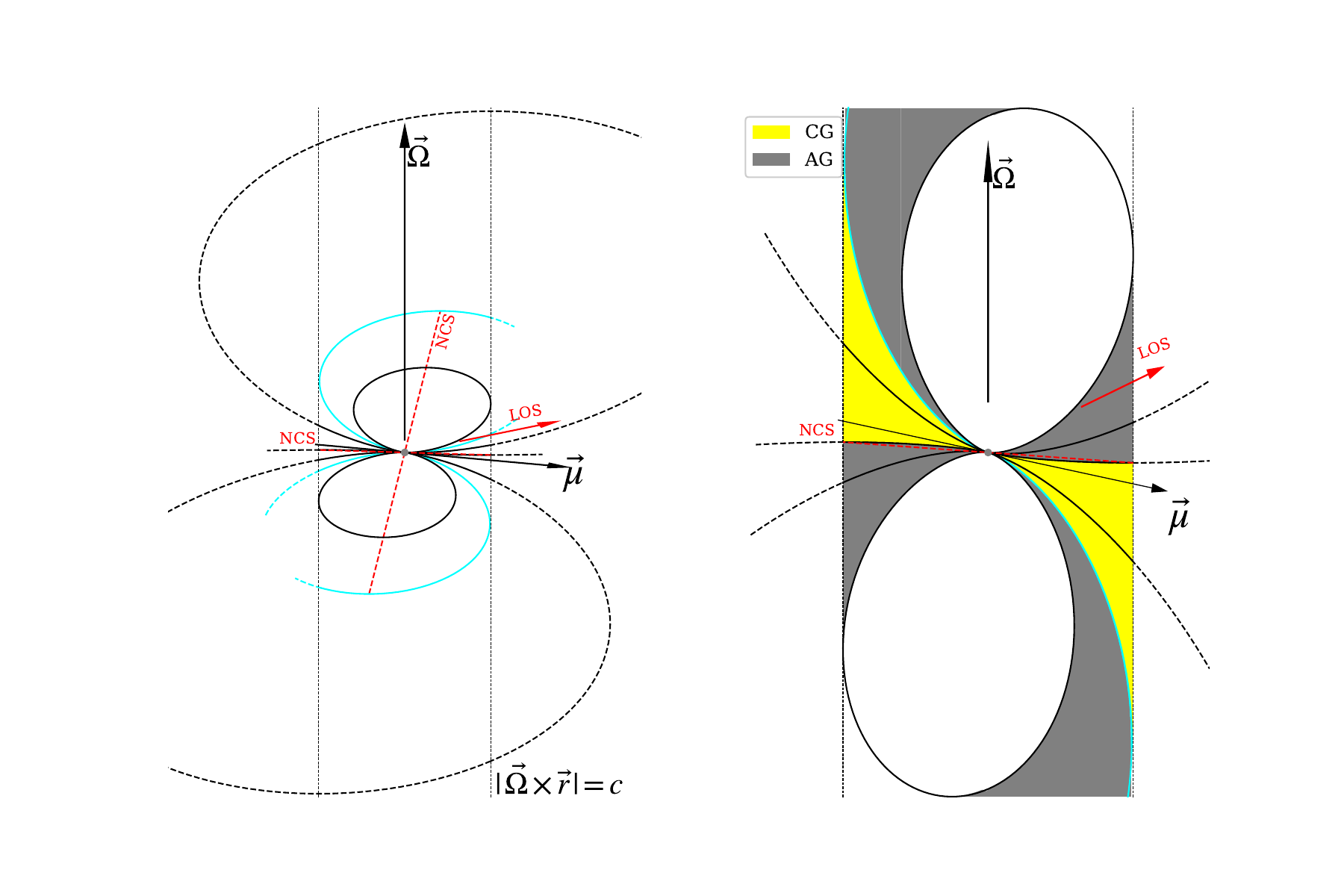}}
 \centerline{(b)}
\end{minipage}
\caption{The radiation geometry of PSR B0950$+$08 using the RVM solutions of $\alpha = 100.5^{\circ}$ and $\beta = -33.2^{\circ}$: (a) three-dimensional magnetospheric geometry of the magnetic dipole field of this pulsar. To understand the magnetospheric geometry, the magnetic pole whose angle with respect to rotation axis equals $ \alpha = 100.5^{\circ} $ is defined as the magnetic Pole 1 in the $ (\mathbf{\Omega} - \mathbf{\mu}) $ plane. Its opposite magnetic pole is defined as the magnetic Pole 2. The last closed field lines are indicated by the cyan curves, and the magnetic field lines originate from the polar cap of the magnetic Pole 1, and Pole 2 are denoted by the blue and red curves, respectively. In addition, the light of sight (LOS) is indicated by the dashed arrow; (b) the magnetospheric dipole geometry of this pulsar in the $ (\mathbf{\Omega} - \mathbf{\mu)} $ plane. The light cylinder radius is indicated by the vertical dotted lines. The light of sight (LOS, red arrow) and the null charged surface (NCS, red dashed line) are also shown, respectively. To reveal the width of the annular gap (AG) and the core gap (CG), the areas between the critical field lines and the last closed field lines depict the AG are filled by the grey, and the CG describes the zones between the magnetic axis and the critical field lines are filled by the yellow in the right-hand panel. The cyan lines depict the magnetic field lines which only cross the light cylinder on the side of the opposite magnetic pole. The magnetic field lines inside and outside the corotating frame of this pulsar are indicated by the solid and dotted lines, respectively. In panel (b), to display clearly the global magnetosphere as well as the polar cap's surroundings, the ratio of the vertical to horizontal dimensions is chosen as $3:1$ for the left-hand panel, while $4:3$ for the right-hand panel.}
\label{geo}
\end{figure*}
\begin{figure}
    \centering
    \includegraphics[width=.8\linewidth]{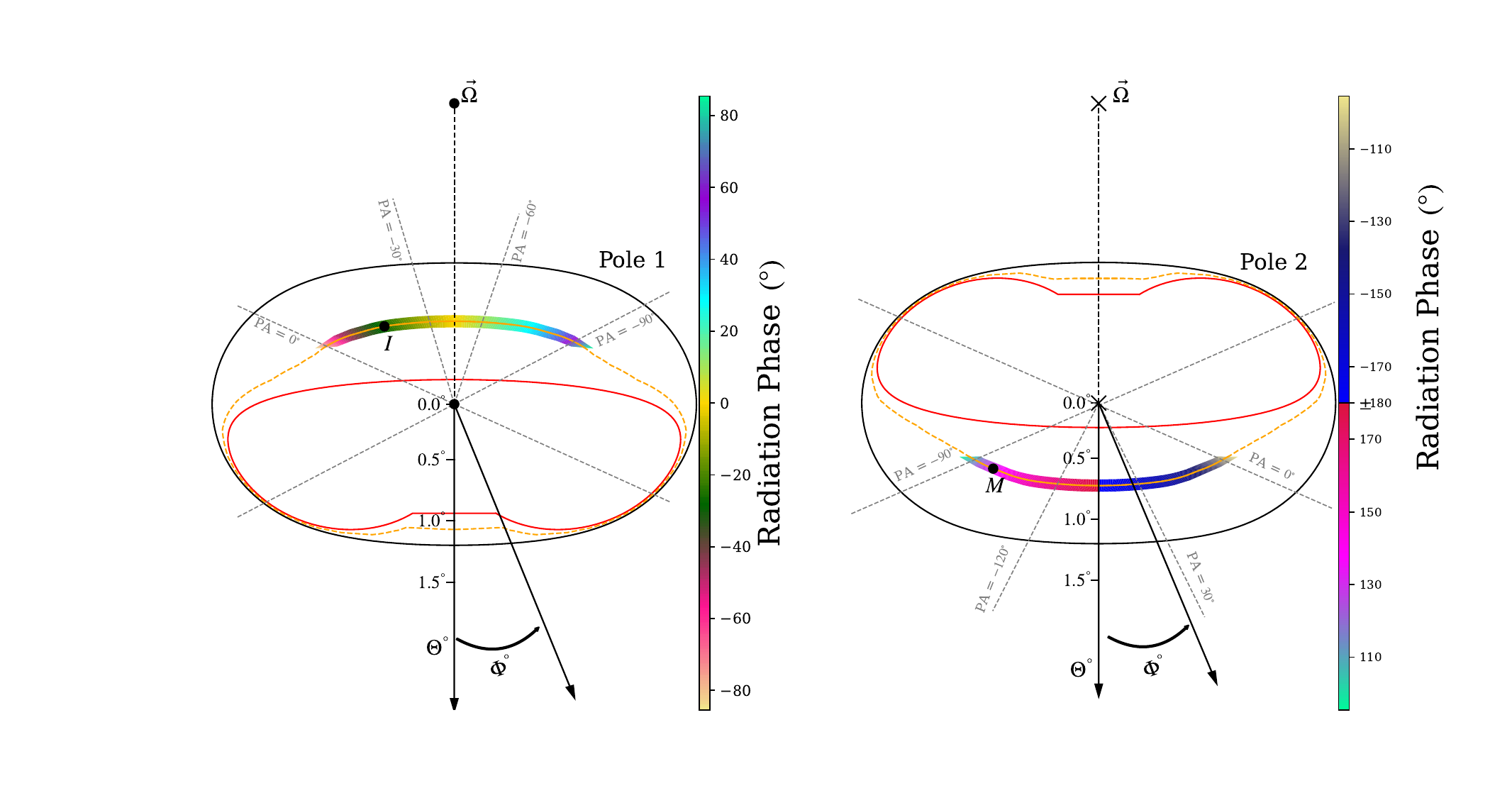}
    \caption{Polar cap shape of PSR B0950$ + $08, the AG and CG of this pulsar are plotted. The black (red) lines represent the intersection curve of the stellar surface and the last closed field lines (the critical field lines). The black dot (x mark) indicates the magnetic pole in the left (right) panel. As shown in the figure, the AG describes the zones between the black and the red curves, the CG is surrounded by the red curve. The radiation of the AG is first used to explain the radio emission of this pulsar. Without loss of generality, we assume that the radio emission of this pulsar emits from these magnetic field lines whose footprints concentrate in the orange line. To unveil the radiation trajectories, the radiation phase is calculated, and the values of the radiation phase are indicated by the colors. The dot denotes the location of the peak of the interpulse (labeled $I$) in the left panel, and it represents the location of the peak of the main pulse (labeled $M$) in the right panel. The magnetic field lines whose footprints concentrate in the orange curve are indicated by the solid line, while their radiation directions are parallel to the telescope inside the corotating frame of this pulsar. The magnetic field lines whose radiation directions are not parallel to the direction of the light of sight, their footprints are denoted by the dashed line. To unravel the emission geometry of this pulsar even further, the PPAs are also plotted (gray dashed lines). The $ \Theta $ and $ \Phi $ denote the polar and azimuthal angles around the magnetic axis, respectively.}
    \label{f5}
\end{figure}
\begin{figure}
    \centering
    \includegraphics[width = .8\linewidth]{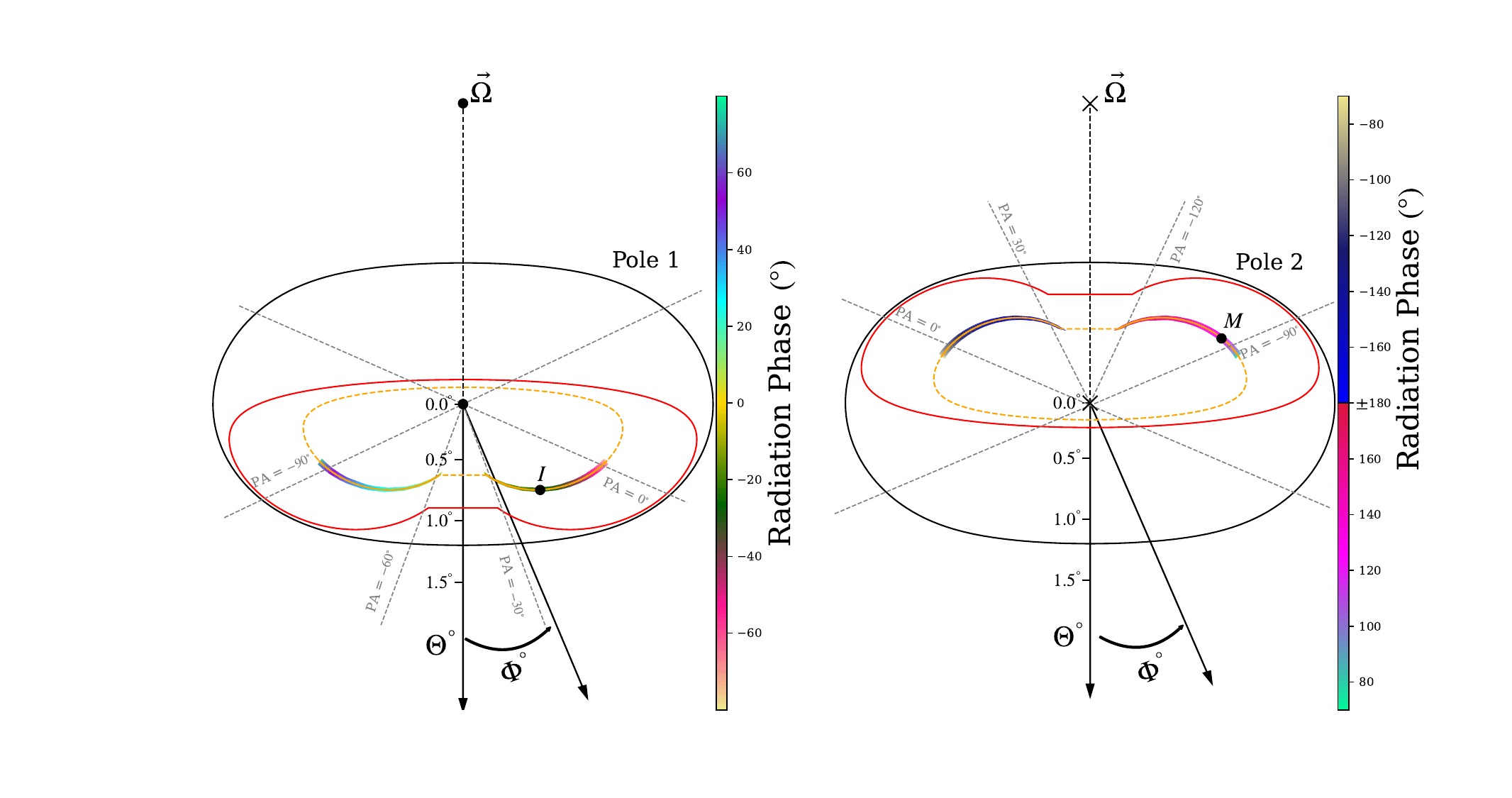}
    \caption{Same as Figure \ref{f5} except for assuming that the radio emission of this pulsar originates from the CG.}
    \label{f6}
\end{figure}
\begin{figure}
    \centering
    \includegraphics[width = .8\linewidth]{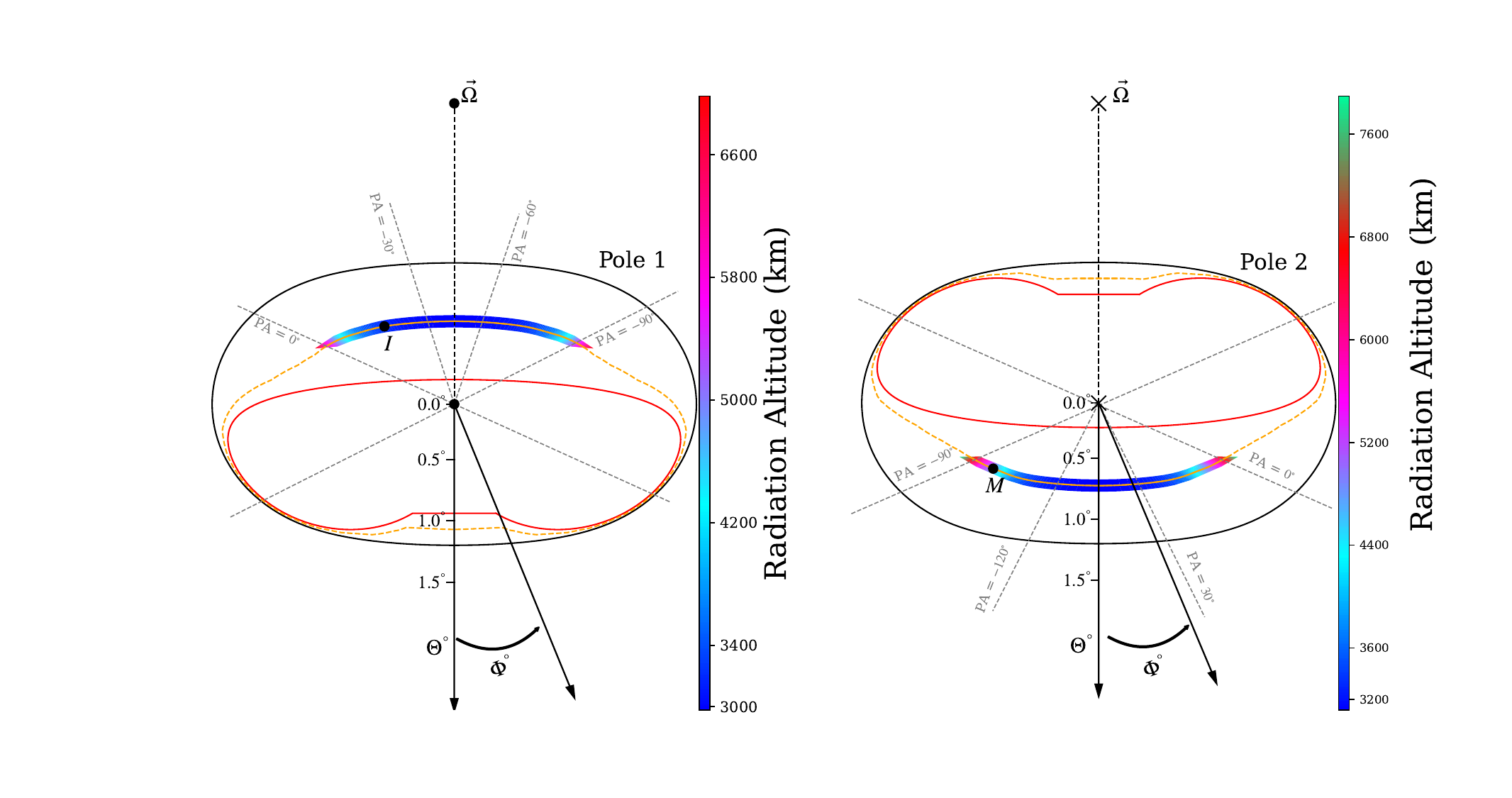}
    \caption{The emission height of this pulsar for the AG model. The left-hand panel indicates the emission heights of the magnetic Pole 1, and the interpulse comes from $\sim 3500$\,km. The altitudes of the magnetic Pole 2 are plotted in the right-hand panel, which shows that the emission height of the main pulse is $\sim 5000$\,km. One can see that the emission height of the main pulse is higher than the interpulse. The bridge component originates from an extremely high altitude $\sim 7800$\,km. The light cylinder radius of this pulsar is ${R_{\rm LC}}$ $\sim 12000$\,km. To further clarify the emission geometry of this pulsar, the PPAs are also included in the plots (gray dashed lines).}
    \label{EH0}
\end{figure}
\begin{figure}
    \centering
    \includegraphics[width = .8\linewidth]{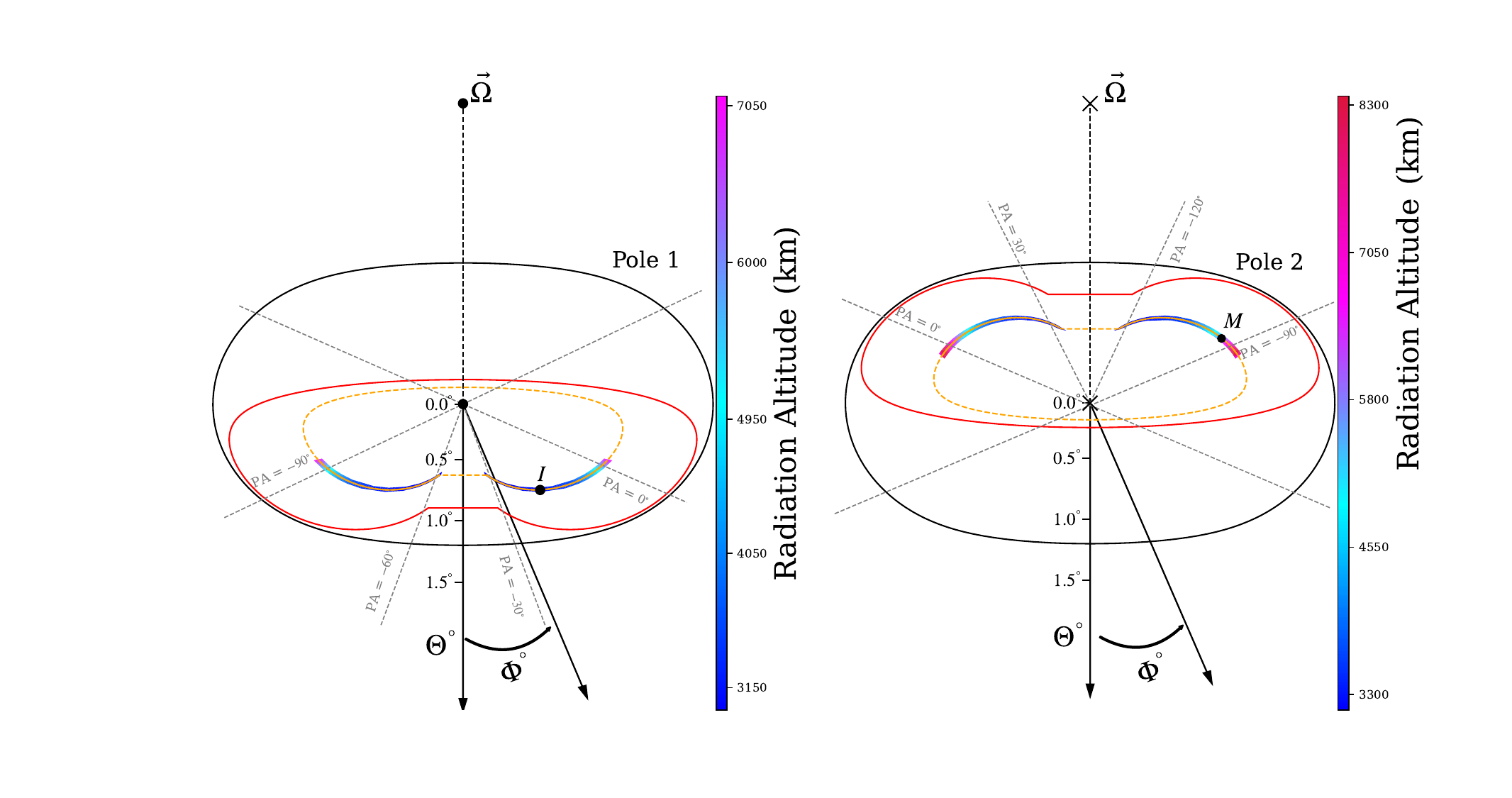}
    \caption{Same as Figure \ref{EH0}, but for the CG model. The emission heights are generally higher, while the CG scenario is considered. In the right-hand panel, it indicates that the emission heights of the weak emission are more than $8000$\,km, and the altitude of the bridge component even reaches $\sim 8500$\,km.}
    \label{EH1}
\end{figure}
%
%
\begin{figure}
    \centering
    \includegraphics[width = .6\linewidth]{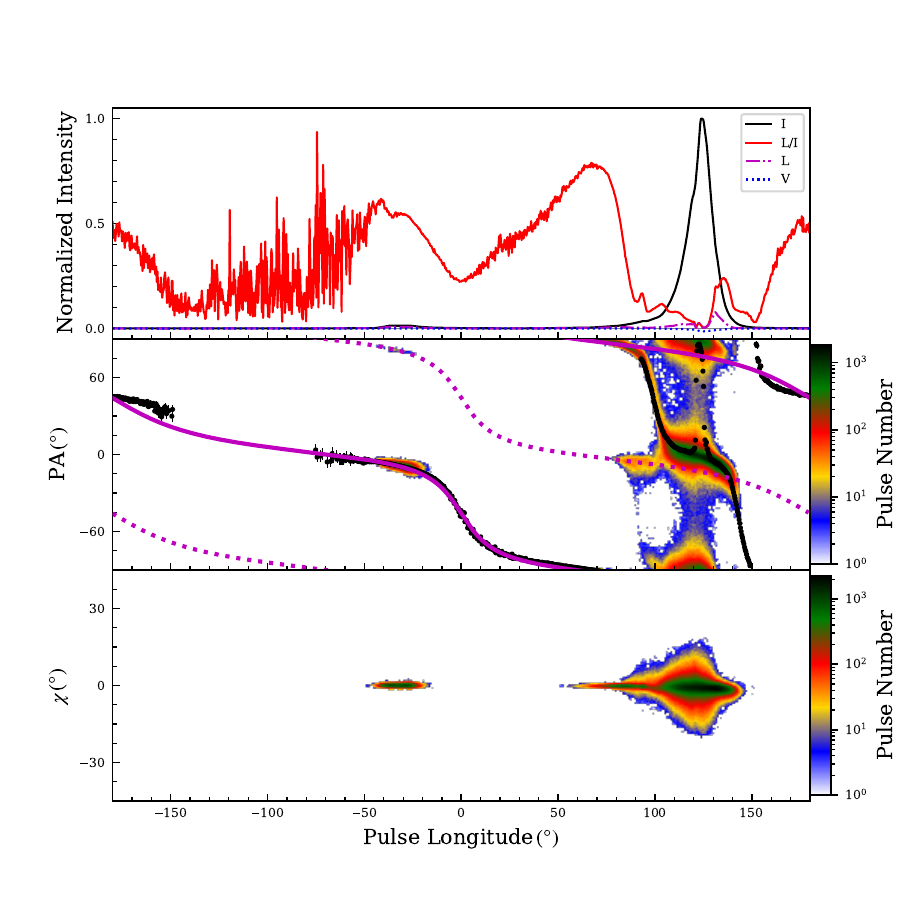}
    \caption{Polarization emission features of the average pulse profile, total radio emission (solid black), linear polarization fractions (solid red), linear polarization (dash-dotted magenta), and circular polarization (dotted blue) are plotted in the top panel. The intensity is scaled with the radio emission peak. The middle panel depicts the distribution of the polarization position angle (PPA) of single pulses as a function of the pulse longitude. The pulse number in each pulse longitude is indicated by the color scale and the average PPA behavior is shown as the black error bars. The best-fitting RVM is denoted as the thick magenta line, and its $90^{\circ}$ offsets are in dotted magenta. It can be found that different polarization modes dominate the polarization emission features of the single pulses in the main pulse. The ellipticity angle $\chi$ distribution of the single pulse is shown in the bottom panel to unravel the circular polarization emission contribution to the radiation in the magnetosphere. We have chosen the error bars for PPA that are less than $5^{\circ}$ for each single pulse. Although such a high threshold reduces the counts and selects against the extremely weak emission pulses, this threshold allows us to distinguish the intrinsic radio emission of the distribution from emission introduced by the system noise.}
    \label{f10}
\end{figure}
\begin{figure}
    \centering
    \includegraphics[width = \linewidth]{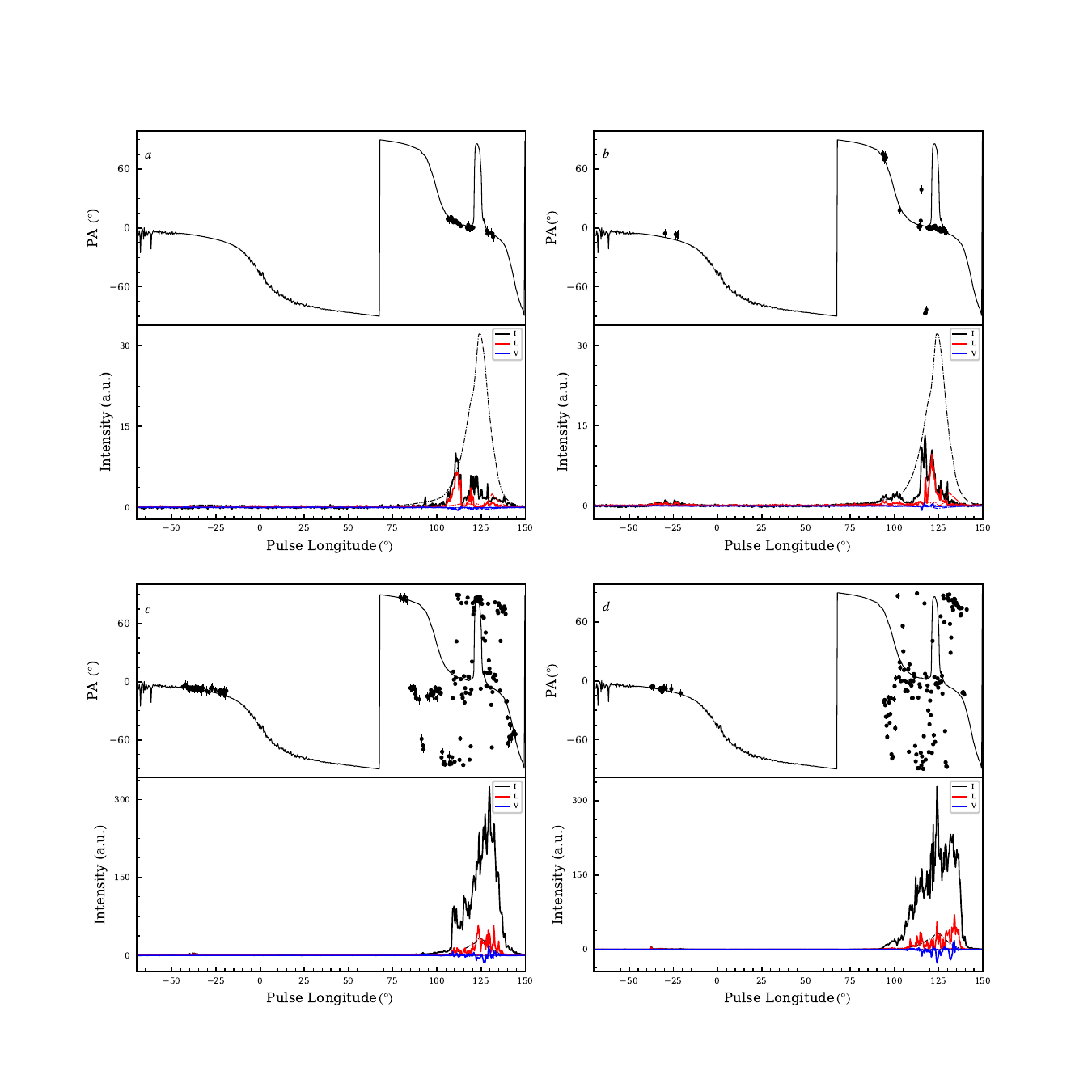}
    \caption{The polarization emission features of the sample of 4 single pulses. As panel $a$ shows, the PPAs of the single pulse are in black dots, and the average PPA track is indicated by the solid black line in the top subpanel. Only the pulse longitudes with the polarization position angle (PPA) error bars less than $5^{\circ}$ for each single pulse are included. The bottom subpanel depicts the Stokes-parameter profiles. The total radio emission intensity, the average linear polarization intensity, and the circular polarization intensity are denoted as dash-dotted black, dash-dotted red, and dash-dotted blue curves, respectively.}
    \label{f11}
\end{figure}
\section{results} \label{sec3}

\subsection{Polarization Emission Property} \label{sub31}

Two tentative baseline intensity determinations for PSR B0950$+$08 described in section \ref{sec2} are proposed, and the conventional baseline subtraction is also taken into account. After subtracting the baseline of this pulsar using three methods, the polarization emission properties of the main and inter pulses are consistent, but the PA-swings of the weak emission region (i.e., the precursor of the main pulse and the postcursor of the interpulse) have little difference. 

The polarization emission properties of this pulsar are shown in the left-hand panels of Figures \ref{f0} and \ref{f1}.
%
It can be found that the polarization emission of the main pulse displays a depolarization phenomenon (the linear fractions are less than $ \sim 10 \%$ ), and there is a significant position angle jump at the location of the peak of the main pulse. Detailed discussions for the physical mechanism of the depolarization and position angle jumps of radio pulsars are presented in \citet{1997A&A...323..395X}, and they concluded that these polarization emission behaviors may be caused by the phase shift of beam centers of the different components of the pulsars.
Other possible origins of the jump such as the orthogonal polarization modes (OPMs) also discussed in detail~\citep[e.g.,][]{2017MNRAS.472.4598D}.
Moreover, the curve of the linear fraction of the main pulse becomes complicated, and the values jump at some pulse longitudes.
To reveal the polarization emission properties of this pulsar, the baseline position is also pointed out with the grey dashed line in the top panels.

\subsection{Fitting polarization position angle with RVM} \label{sub32}
To understand the magnetospheric geometry of PSR B0950$ + $08, the classical RVM is used to fit the observed PPAs of this pulsar \citep{1969ApL.....3..225R}. The PPAs of the pulse longitudes whose linear fractions are higher than $ \sim 30 \% $ are considered in the fit, and more discussions will be summarized in section \ref{sec5}. After taking the position of the steepest gradient of the RVM fit curve as the reference pulse longitude $0^{\circ}$, the results are shown in Figure \ref{f0}. 
It can be found that the steepest gradient of the RVM fit curve (the reference pulse longitude $ 0^{\circ} $) approaches the interpulse. This property demonstrates that the distance between the magnetic pole and the interpulse is closer than the main pulse.

In the $ \alpha - \zeta $ panel, the values of $ \chi_{\mathrm{reduced}}^2 $ are indicated by the colors. The cross denotes the minimum value of \textbf{$ \chi_{\mathrm{reduced}}^2 $} in the fit, and it gives the values of the viewing angle $ \zeta = 67.3^{\circ}$ and the inclination angle $ \alpha = 100.5^{\circ}$. Compared with this result, the baseline of this pulsar is subtracted using the conventional baseline subtraction, and the results are shown in Figure \ref{f1}.
In the right-hand panel, it indicates the viewing angle $\zeta = 59.4^{\circ}$ and the inclination angle $\alpha = 96.5^{\circ}$.
The region between the two vertical blue lines is believed to be the baseline region since it is far away from the main pulse. Compared with other pulse longitudes, the impulsive radio emission detected in the chosen region is extremely weak and close to the baseline intensity. This region is regarded as the reference of the baseline and subtracting it minimizes the effect of the baseline subtraction on the pulsar with emission features of the whole $360^{\circ}$ of longitude~\citep{2022MNRAS.517.5560W}.
One can see that the values of the $ \chi_{\mathrm{reduced}}^2 $ of this RVM solution of $\alpha$ and $\zeta$ are about ten times as much as the values of $ \chi_{\mathrm{reduced}}^2 $ of the results shown in the right-hand panel of Figure \ref{f0}, and the distribution of the $\alpha$ and $\zeta$ of this RVM solution is wider, which means larger uncertainty. For these reasons, the $\alpha = 100.5^{\circ}$ and $\zeta = 67.3^{\circ}$ are used for the later analyses of this work. The RVM solutions of the $\alpha $ and $\zeta$ given in Figures \ref{f0} and \ref{f1} are consistent with each other. ~\citet{2024ApJ...963...65W}

\subsection{The Radiative Geometry in Magnetosphere} \label{sub33}
The magnetospheric geometry of PSR B0950$+$08 is revealed by assuming the magnetic dipole field. In Figure \ref{geo}, panel (a) depicts the three-dimensional magnetospheric geometry of this pulsar. Panel (b) describes the magnetospheric geometry of this pulsar in the ($ \mathbf{\Omega} - \mathbf{\mu} $) plane, and the light cylinder radius of this pulsar is represented by the vertical dotted lines. Its rotation axis is upward, and the magnetic axis is inclined with an angle of $ \alpha $ in the ($ \mathbf{\Omega} - \mathbf{\mu} $) plane. The null charged surface is the surface where the magnetic field line is perpendicular to the rotation axis (i.e., $\mathbf{\Omega} \cdot \mathbf{B} = 0$) is indicated by the red dotted lines (labeled NCS). Detailed descriptions of these figures are given in their captions.

To understand the $ \gamma $-ray and radio emission from the pulsar, \citet{2004ApJ...606L..49Q} proposed a new segment of the polar cap of the pulsar~\citep[][]{1975ApJ...196...51R}. The conventional RS75-type gap model is segmented as the annular gap (AG) and core gap (CG), and the boundary between the AG and the CG is defined by the magnetic field lines intersecting the null charge surface (NCS). The AG describes the region between the last closed field lines and the critical field lines (the magnetic field lines are perpendicular to the rotation axis (i.e., $\mathbf{\Omega \cdot B} = 0$) at the light cylinder), the CG is surrounded by the critical field lines.

In the right-hand panel of Figure \ref{geo}, the AG and CG are also plotted. To unveil the width of the two gaps, the AG and CG are filled with grey and yellow colors, respectively. In the polar cap between the magnetic and rotation axes, the width of the CG is extremely narrow compared to the width of the AG. Moreover, the width of the CG is also narrower than the width between the equator and the magnetic axis since the equator falls in the grey regions in the magnetic Pole 1. One can see that the last closed field lines become more elliptical. For the pulsar whose inclination angle $\alpha < 90^{\circ}$, as pointed out by \citet{2004ApJ...606L..49Q}, the width of the AG between the magnetic axis and the equator becomes wider when $ \alpha $ increases. On the contrary, the width of the AG between the magnetic and rotation axes becomes narrow. The width of the AG of PSR B0950$+$08 is different with $ \alpha < 90^{\circ} $ scenarios \citep{2004ApJ...606L..49Q}. As panel (b) of Figure \ref{geo} shows, the equator falls in the region between the magnetic and rotation axes. The width of the AG between the magnetic axis and equator (or rotation axis) becomes wider, but the width of the AG between the magnetic axis and the anti-parallel direction of the rotation axis becomes narrow. In magnetic Pole 2, the variation of the width of the AG is similar to the scenarios $\alpha < 90^{\circ}$ discussed in \citet{2004ApJ...606L..49Q}. 

In the AG and CG models, the potential drop of the polar cap of the pulsar is related to the width of the AG. A larger width of the AG results in a higher potential drop, making it easier to produce sparks. Based on this physical picture, in magnetic Pole 1, the opening field line region between the magnetic and rotation axes is easier to produce sparks than the region between the magnetic and the anti-parallel direction of the rotation axes. On the contrary, in magnetic Pole 2, the region between the magnetic axis and the equator is easier to produce sparks than the opening field line region between the magnetic and rotation axes.

It can be seen that the magnetospheric structure of this pulsar becomes complex since a large inclination angle $\alpha$.
As panel (b) of Figure \ref{geo} shows, the cyan field lines only across the light cylinder on the side of the opposite magnetic pole.
This structure implies that some of the opening field line regions in which the magnetic field lines are not passing through the light cylinder.

\subsection{Emission Zones in both Regions of AG and CG} \label{sub34}
We unravel the polar cap surface of PSR B0950$+$08 based on the 110-min polarization observation under the framework of the dipole field. To understand which opening field line region dominates this pulsar's radiation, the AG and CG of the vacuum gap model are used to reveal its emission geometry.
The shapes of the AG and the CG of the dipole field of PSR B0950$ + $08 at the stellar surface are shown in Figures \ref{f5} and \ref{f6}, and two-pole model scenarios are plotted.
A detailed description of the emission zone is given in the caption of the figure.

One can see that the equator of this pulsar falls in the region between magnetic and rotation axes, this structure is different with $ \alpha < 90^{\circ} $ scenarios discussed in \citet{2004ApJ...606L..49Q}. \citet{2004ApJ...606L..49Q} considered the magnetospheric geometry of the magnetic dipole field of the pulsars for different inclination angles $\alpha$ (i.e., $ \alpha = 0^{\circ}, 45^{\circ}, 75^{\circ}$), and they found that the width of the AG is a function of the inclination angle $\alpha$. As Figure \ref{f5} shows, in the left-hand panel it reveals that the width of the AG between the magnetic and rotation axes is wider than the width of the AG between the magnetic axis and the anti-parallel direction of the rotation axis. On the contrary, from the right-hand panel, the variation of the width of the AG is similar to \cite{2004ApJ...606L..49Q} since the value of the inclination angle is $ \alpha = 79.5^{\circ} $ in the magnetic Pole 2. As the panel (b) of Figure \ref{geo} shows, the width of the AG between the magnetic axis and the anti-parallel direction of the rotation axis becomes narrow, because the critical field lines that fall in the region between the magnetic axis and the cyan magnetic field lines are perpendicular to the rotation axis (i.e., $\mathbf{\Omega} \cdot \mathbf{B} = 0$) only when they intersect with opposite side light cylinder in the ($ \mathbf{\Omega} - \mathbf{\mu} $) plane.

\subsection{A Two-pole Model of PSR B0950$+$08}

The charged particles will emit radiation, while they move along the magnetic field lines whose intersection curves with the stellar surface are inside the polar cap of the pulsar \citep[e.g.][]{1975ApJ...196...51R,1979ApJ...231..854A}. The radiation of the charged particles will be detected by the telescope, while their radiation directions are parallel to the light of sight. The magnetic field lines in the corotating frame of the pulsar have no more than one point where the radiation direction is parallel to the light of sight before they intersect with the light cylinder. Therefore, the radiation phase of the emission points whose radiation directions are parallel to the direction of the light of sight can be determined. 

Both the AG and the CG are taken into account to understand the radio emission features from PSR B0950+08. As Figure \ref{f5} shows, without loss of generality, the orange curve obtained by averaging the distances between the last closed field lines (black curve) and the critical field lines (red curve) is used to denote the radiation generated from the AG. According to the result of the calculations, the radio emission of this pulsar originates from the magnetic Pole 1 shown in the left-hand panel, and the radiation generated from its opposite magnetic pole is described in the right-hand panel. 
When the radiation direction of the magnetic field lines is parallel to the telescope, the magnetic field lines whose footprints are concentrated on the orange curve are indicated by the solid line. Meanwhile, the CG scenario is shown in Figure \ref{f6}, where the two-thirds distances between the critical field lines (red curve) and the magnetic pole are used to indicate the radiation of the CG. 

According to the radio emission properties of PSR B0950$+$08 \citep{2022MNRAS.517.5560W} and its magnetospheric geometry based on the polarization-calibrated observation in this work, the emission points whose radiation directions are parallel to the direction of the light of sight can be determined using the geometric relation $\zeta = \cos^{-1} (\mathbf{\hat{B}_z}/ \sqrt{\mathbf{\hat{B}_x}^2 + \mathbf{\hat{B}_y}^2 + \mathbf{\hat{B}_z}^2})$, where $\mathbf{\hat{B}} = \mathbf{\hat{B}_x} + \mathbf{\hat{B}_y} + \mathbf{\hat{B}_z}$ is the unit vector of the magnetic field of the emission point. To exhibit the radiation trajectories of this pulsar, the radiation phase is calculated, and the values of the radiation phase are indicated by the colors. After taking the steepest gradient of the RVM fit curve as the reference pulse longitude $0^{\circ}$, Figure \ref{f0} indicates that the locations of the peak of the interpulse and main pulse are $-25^{\circ}$ and $125^{\circ}$, respectively. As Figure \ref{f5} shows, the locations of the peak of the interpulse and main pulse are also denoted. One can see that the radio emission of the pulse longitudes from $ \sim -85 $ to $ \sim 85^{\circ}$ of this pulsar originates from the magnetic Pole 1, while the radio emission of the left $52\%$ of the rotation period comes from its opposite magnetic Pole 2. As Figure \ref{f0} shows, the polarization emission properties of this pulsar originate from the magnetic Pole 1 and can be described by the RVM curve well. Meanwhile, the radio emission of the main pulse comes from the opposite magnetic Pole 2. The polarization emission properties of the main pulse display the depolarization and position angle jump behaviors. These polarization phenomena imply that the polarization emission originates from the magnetic Pole 2 significantly deviates from RVM.

Figure \ref{f6} depicts that the radiation is generated from the CG.
Similar to the radiation of the AG scenario, we assume that the magnetic field lines whose footprints concentrate in the orange curve indicate the radiation of the CG. It can be seen that the radiation of the pulse longitudes from $\sim -75 $ to $ \sim 75^{\circ}$ originate from the magnetic Pole 1, and the radiation of the left $58 \%$ of the rotation period of PSR B0950$+$08 come from the magnetic Pole 2.

The phenomenological model used to explain the radio emission properties of PSR B0950$+$08 would provide further insight into the emission zone, but for the actual scenario, the magnetic field lines intersecting with the stellar surface would become more complex. According to the whole pulse phase radiation characteristics and polarization emission features, the calculations of the radiation of the AG and the CG models indicate that the emission of the interpulse comes from the magnetic Pole 1, while the radiation of the main pulse originates from the magnetic Pole 2.

\subsection{The Emission Height}

Based on the result of the mapping sparks on the polar cap surface shown in Figures \ref{f5} and \ref{f6}, the emission height can be calculated according to the three-dimensional calculations of the pulsar emission height given in~\citet{2007A&A...465..525Z}.
With the AG model, the height of the emission point can be calculated. The results are shown in Figure \ref{EH0}, and the emission heights are indicated by the colors. In the left-hand panel, it shows the emission height of the magnetic Pole 1, and the altitudes range from $\sim 3000$ to $\sim 7000$\,km. In addition, the emission heights of the radiation phase which originates from the magnetic Pole 2 are shown in the right-hand panel, which indicates that the heights range from $\sim 3200$ to $\sim 7800$\,km. It can be seen that the emission heights of the interpulse and main pulse are $\sim 3500$\,km and $\sim 5000$\,km, respectively. Moreover, the bridge component emits from a higher altitude of around $7800$\,km.
One can find that the variation in the emission height is related to the distance between the emission regions and the magnetic pole. The calculation indicates that the emission height of the region that approaches the magnetic pole changes slowly. Meanwhile, the emission height changes sharply, while the emission zone is away from the magnetic pole.

With the CG scenario, the emission height is also calculated, and the results are shown in Figure \ref{EH1}. One can see that the variation of the emission height is similar to the AG scenario, but the emission heights of the CG are generally higher than the AG. Furthermore, it can be found that the emission height of the bridge component even reaches $\sim 8500$\,km.

The analysis of the AG and the CG models indicates that PSR B0950$+$08 is a high-altitude magnetospheric emission pulsar. High radiation altitudes challenge the conventional particle acceleration mechanism and coherent emission mechanism of the radio pulsars \citep[][]{1975ApJ...196...51R}. With the high radiation altitudes and the whole pulse phase radiation characteristics, PSR B0950$+$08 may have an additional accelerating electric field such as the slot gap accelerator \citep[][]{1983ApJ...266..215A} or annular gap accelerator \citep[][]{2007ChJAA...7..496Q} to accelerate the charged particles.

\section{Discussions}
\label{sec5}

%

The pulsar's surface physics is hard to probe due to the absence of direct observations. Compared with other normal radio pulsars, the emission geometry of PSR B0950$+$08 would be good for unraveling the electrodynamics of the pulsar magnetosphere related to the stellar surface.
%
%
Recently, several research groups published polarimetric emission characteristics of radio pulsars.
\citet{2023MNRAS.520.4801J} reported the application of the RVM to a sample of 854 radio pulsars observed with the MeerKAT telescope.
They argued that the average PPAs tracks of $60 \%$ of the pulsars can be described by the RVM. For the pulsars with good RVM fits, their polarization emission features are similar to the Vela pulsar with a high fraction of linear polarization.
%
\citet{2023MNRAS.524.5042R} presented the polarization emission properties for the 60 radio pulsars using the Arecibo radio telescope.
They did not present the RVM fits for these objects, but the variation of the PPAs of these pulsars can be revealed by analyzing their Stokes-parameter profiles.
It can be found that the polarization behaviors of the pulsars with visible linear polarization fractions in the emission regions are also similar to those of the Vela pulsar. Moreover,~\citet{2023RAA....23j4002W} published the polarization profiles of 682 pulsars using the FAST radio telescope.
For the pulsars with a high fraction of linear polarization, the steepest gradients of the RVM-fitted curves are almost aligned with the radio emission peaks, implying that the sparking on the polar cap surface for these pulsars is regular and symmetrical.

\par The emission geometry of PSR B0950$+$08 would be applied to understand the radiation in the magnetosphere and the sparking on the polar cap surface for the normal radio pulsars whose emission occupies almost the whole pulse phase (e.g., PSRs J1851$+$0418, J1903$+$0925, J1916$+$0748 and J1932$+$1059~\citep[e.g.,][]{2023RAA....23j4002W}).
In the future, we will continue to investigate the emission geometries of the four normal pulsars with extremely wide profiles using the FAST radio telescope.
The number of pulsars that exhibit radio emission peaks misaligned with their magnetic poles after accurately mapping the sparks on the polar cap surfaces and whether these pulsars occupy a specific region in the $P-\dot{P}$ diagram are questions that need to be further studied.
It is worth investigating whether the emission geometries of these objects with extremely wide profiles can better reflect the physics of their polar cap surfaces.
The FAST radio telescope will play a key role in unraveling the stellar surface and even in understanding the radiation mechanism of the radio pulsar, which is our interest now.

Up to now, the offsets of the point of highest PA slope with respect to radio emission peak are small compared to that of PSR B0950$+$08~\citep[e.g.,][]{2023MNRAS.520.4801J,2023RAA....23j4002W}.
The offset for PSR B0950$+$08 is approximately equal to $60^{\circ}$ under the framework of the two-pole model.
Most of the radio pulsars whose offsets are less than $25^{\circ}$, and that of few pulsars reach $45^{\circ}.$ 
Small offsets are understandable, explained by the underlying physics of rotation-induced relativistic effects (i.e., aberration and retardation), which do not reflect the electrodynamics of the magnetosphere related to stellar surface physics.
The maximum offset between the point of the highest PA slope and the radio emission peak is found based on the 110-min polarization observation with the FAST radio telescope. After mapping the sparking points on the surface (see Figures~\ref{f5} and~\ref{f6}), the polar cap sparking pattern of PSR B0950$+$08 implies that the sparks distribution is non-symmetrical and irregular.
This finding would bring new ideas into the pulsar's surface physics and even the inner dense matter of the pulsar.
A pulsar's magnetospheric plasma and its inner dense matter are separated by the stellar surface, and the radiative mechanism could thus be essentially a problem relevant to the nature of the pulsar's surface and even of cold matter at supra-nuclear density.
This relationship could be leaked by the polarization observation of PSR B0950$+$08: Gaussian-like means pulse profiles would usually peak at the maximum slopes of PPA sweep~\citep[e.g., the Vela pulsar;][]{1969ApL.....3..225R}, however, both the main and inter pulses from two poles are not at the maximum of PSR B0950$+$08.
Vela-like polarization could be understandable if sparking points distribute regularly along either AG or CG trajectories on the polar cap and the consequent pair-plasma moves along a flux tube,\footnote{%
The radiation of a bunch of charged particles could be coherent in case of ordering of condensation in either momentum space (maser) or position space (antenna). A spark would be able to achieve simply the antenna mechanism for coherence which should become weaker as the dispersing plasma moves outer.
%
Such a sparking point may result in a fan-leaf of emission, the evidence for which could have already been provided observationally~\citep{2014ApJ...789...73W,2019Sci...365.1013D}.
} forming a fan-shaped pattern~\citep{2014ApJ...789...73W}, however, as illustrated in Figs.~5-8, the mapped sparks on PSR B0950$ + $08's surface are in-homogeneously distributed away from the magnetic pole.
This may hint preferential discharges somewhere on the polar cap due to a rough pulsar surface (there are protuberances (i.e., small mountains) on the stellar surface that would make its surface rough), as had already been proposed for understanding FAST's single pulse observation of PSR B2016+28~\citep{2019SCPMA..6259505L}.
%
%
Regular drifting sparks are speculated to occur for Vela-like young and energetic pulsars, while sparse sparks are for old and lazy ones whose magnetospheric activity is weak and thus not so diligent at work.

In what kind of pulsar inner structure model could a stable rough surface be reproduced?
The answer to this question is surely related to Landau's ``giant nucleus'' anticipated superficially more than 90 years ago~\citep[e.g.,][]{2013PhyU...56..289Y,2023AN....34430008X}, and we are faced now with choices~\citep[e.g.,][]{1999ApJ...522L.109X} of either conventional neutron star or strange quark star, and even of strangeon star.
In this sense, a comprehensive study of radio pulsar emission with highly-sensitive FAST is admittedly meaningful for the physics of dense matter.

Observation of polarized emission is necessary to identify pulsar radiative geometry, the polarization features, however, depend on the ways of baseline subtraction.
%
For a single-dish radio telescope, it is difficult to determine accurately the baseline position of PSR B0950$+$08 since its radio signal is detected over the whole pulse phase \citep{2022MNRAS.517.5560W}.
About those two tentative baseline intensity determinations, the left-hand panels of Figures \ref{f0} and \ref{f1} indicate that the PA swings in the precursor of the main pulse and the postcursor of the interpulse are different.
Compared with the left-hand panel of Figure \ref{f0}, we find that the RVM fit curve shown in the left-hand panel of Figure \ref{f1} becomes flatter.
Nevertheless, both RVM solutions are similar, showing that the interpulse originates from the magnetic Pole 1, while the main
pulse from the Pole 2.
As shown in Figures \ref{EH0} and \ref{EH1}, the emission height of this pulsar is extremely high.
Compared with the CG model, one may suggest that the AG model would be a better one to understand the radio emission of PSR B0950$+$08 since the radiation altitudes of the CG are generally higher than that of the AG.
The bridge component could even come from $\sim 8500$\,km if the CG model is applied to explain the radiation.
Several relativistic effects such as aberration, retardation, and the magnetic field lines sweep back~\citep[e.g.,][]{2006MNRAS.366..945W} should be included in mapping the sparking points on the polar cap surface. When these relativistic effects are taken into account, the sparking points on the polar cap would have slightly adjusted. However, the finding that the strong emission regions (i.e., main and inter pulses) of this pulsar are away from its magnetic pole is hardly changed since the magnetic field lines sweep back causing the emission from lower radiation altitudes earlier than that of from higher, which is the opposite of the aberration and retardation effects~\citep[e.g.,][]{2006MNRAS.366..945W}. In the future, we will consider these effects when mapping the sparking points on the polar cap surface to obtain an accurate polar cap sparking pattern.

\citet{1984ApJS...55..247S} presented single pulse polarization observation of PSR B0950$+$08 at 1404\,MHz with the Arecibo radio telescope.
As the result of the polarization distribution display for PSR B0950$+$08 of Figure 19~\citep[plotted in][]{1984ApJS...55..247S} shows, the PPAs of the single pulses in the main pulse are also complex rather than an ``S''-shape.
Meanwhile, the PPAs of the single pulse in the main pulse display frequently position angle jumps at some pulse longitudes, exhibiting the position angle discontinuity.
Moreover, the linear polarization emission features of the single pulse in the main pulse are similar to that of the average pulse profile, whose linear fractions are even lower than $10\%$. The polarization emission features of the single pulses of this pulsar in the main pulse were published in~\citet{1984ApJS...55..247S}, which also justifies our efforts to obtain the RVM geometry of this pulsar.
As Figure~\ref{f10} shows, the PPAs of the single pulses also display discontinuity rather than monotonic. Meanwhile, the depolarization that occurs in the main pulse may be due to the emission consisting of roughly equal contributions from two modes simultaneously.
In addition, as Figure~\ref{f11} shows, the linear polarization fraction of the strong single pulse compared to the average pulse displays extreme depolarization, which is similar to that of the average pulse profile. It is evident that the polarization emission state of these strong single pulses exhibits irregular and random features, while only that of the weak single pulse with a high linear fraction in the main pulse is amenable to the average PPA track. The situation is similar for many other pulses: strong pulses have both orthogonal modes more contributed, making linear fractions lower and leading to PPA curves more complicated and unlikely following the rotating vector model.
%

It is, however, still worth noticing that in the longitude range (from $90^{\circ}$ to $100^{\circ}$), the PPA curve of the average pulse profile is steep and looks like S-shape (in Figures~\ref{f0} and \ref{f10}). Also, the single pulses' PPAs in the longitude range (from $75^{\circ}$ to $100^{\circ}$) seem to be another pair of OPMs, with PPA patches separated from the main pulse's PPA patches. The average PPA track at the longitude range (from $90^{\circ}$ to $100^{\circ}$), along with single pulses' PPA distributed closely around the curve, connects two PPA patches from different pairs of OPMs. Given the discussion above and the fact that one RVM curve could not fit PPAs of both O and X modes at the same time, RVM does not work in the longitude range (from $90^{\circ}$ to $100^{\circ}$) for the average pulse profile's PPA track. That part of PPAs may indicate a conversion in OPMs, such as the linear coupling between O and X modes discussed in \citep{2001A&A...378..883P}, which happens when the two modes become indistinguishable.

%
Compared to the pulse longitudes used in the RVM fit, the main pulse also exhibits a visible contribution of circular polarization, with circular fractions that exceed several times those of the chosen pulse longitudes. This means that circular polarization in the main pulse plays a significant role in the radiation within the magnetosphere.
As pointed out by \citet{1969ApL.....3..225R}, the RVM geometry is based on the variation of the orientation of the magnetic field lines with respect to the ($ \mathbf{\Omega} - \mathbf{\mu} $) plane (e.g., the linear polarization emission contribution to the radiation in the magnetosphere).
This model predicts a monotonic position angle rotation across the pulse profile.
Consequently, the observed PPAs of both the single pulse and average pulse in the main pulse deviate from the frame of the RVM geometry due to these polarization emission features. 
Therefore, the higher linear polarization emission contribution can reflect the RVM geometry well for this pulsar.
In this work, the pulse longitudes whose linear fractions are higher than $\sim 30\%$ are chosen, and the average PPA of the chosen pulse longitudes can be fitted by the RVM curve well.
%

\section{Conclusions} 
\label{sec6}

The determination of baseline is one of the key issues in detecting the radio emission polarization of PSR B0950$+$08 \citep{2022MNRAS.517.5560W}.
In this work, two tentative methods as noted in Equations (\ref{eq0}) and (\ref{eq1}) are proposed to determine the baseline intensity of this pulsar.
After subtracting the baseline intensity with Equations (\ref{eq0}), we present the results shown in Figure \ref{f0}.
In the right-hand panel of Figure \ref{f0}, it indicates the inclination angle $\alpha = 100.5^{\circ}$ and the impact angle $\beta = -33.2^{\circ}$.
The conventional baseline subtraction is also considered, in which the intensity near pulse longitude $-100^{\circ}$ is chosen as the baseline.
The results are shown in Figure \ref{f1}, which implies the RVM solutions of the inclination angle $\alpha = 96.5^{\circ}$ and the impact angle $\beta = -37.1^{\circ}$.
Figures \ref{f0} and \ref{f1} demonstrate that the minimum value of the $\chi_{\mathrm{reduced}}^2$ (the location of the cross) implies similar RVM solutions of the $\alpha$ and $\beta$, though the set of \{$\alpha = 100.5^{\circ}$, $\beta = -33.2^{\circ}$\} would be more acceptable than the other one due to its low $\chi_{\mathrm{reduced}}^{2}$-value about ten times smaller.
Moreover, the distribution of the $\alpha$- and $\zeta$-values becomes compact in Figure \ref{f0}.
Therefore, the tentative methods described in Equations (\ref{eq0}) and (\ref{eq1}) would be suitable for determining the baseline of this pulsar characterized by the whole $360^{\circ}$ of longitude radiation.

Figures \ref{f0} and \ref{f1} indicate also that the polarization emission properties of this pulsar are complex.
The main pulse displays depolarization and position angle jump behaviors, as well as linear fraction variation at some pulse longitudes.
To obtain the RVM solutions of the inclination angle $\alpha$ and the viewing angle $\zeta$ of this pulsar, the pulse longitude ranges from $ \sim 82$ to $152^{\circ}$ and from $ \sim -160$ to $-100^{\circ}$ are unweighted in the fit because of low linear polarization fraction and thus large uncertainty of PPAs.
%

Assuming a magnetic dipole configuration, we could then present the magnetospheric geometry of PSR B0950$+$08.
The three-dimensional view of this pulsar is shown in panel (a) of Figure \ref{geo}.
It is found that the curvature radius of the field lines near the magnetic pole becomes very large.
The panel (b) of Figure \ref{geo} demonstrates CG (yellow) and AG (grey) regions.
%
In the magnetic dipole configuration, the sparking trajectory responsible for the radio emission of PSR B0950$+$08 is investigated in the inner vacuum gap model, for both cases of AG and CG, as depicted in Figures \ref{f5} and \ref{f6}.
It can be found that the radio emission peak is far away from its magnetic pole.
%
%
%
With the AG and the CG models, the calculated emission heights are shown in Figures \ref{EH0} and \ref{EH1}.
It shows that PSR B0950$+$08 is a high-altitude magnetospheric emission pulsar, radiating from heights from $\sim 0.25R_{\rm LC}$ to $\sim 0.65R_{\rm LC}$ (the light cylinder radius $R_{\rm LC}\sim 12000$ km), and even $\sim 7800$\,km for the bridge component.
%

%
\begin{acknowledgments}
{\em Acknowledgments.}  This work made use of the data from the Five-hundred-meter Aperture Spherical radio Telescope (FAST). FAST is a Chinese national mega-science facility, operated by the National Astronomical Observatories, Chinese Academy of Sciences.
This work is supported by the National
SKA Program of China (2020SKA0120100), the National Natural Science Foundation of China (Grant Nos. 12003047 and 12133003), and the Strategic Priority Research Program of the Chinese Academy of Sciences, Grant No. XDB0550300.
\end{acknowledgments}

%






\appendix

\section{The determination of the baseline intensity of PSR B0950$+$08}
\label{secA}

Considering that the pulse profile of the single pulse only occupies a narrow pulse longitude, this emission feature can be used to analyze the baseline response with time for the pulsar with the whole $360^{\circ}$ of longitude radiation characteristics (like PSR B0950+08). Consequently, compared with the average pulse, the baseline intensity in a single pulse is easily determined. The baseline intensity response with time can be determined by analyzing the baseline position in each single pulse. After eliminating the effect of the baseline response with time, the baseline intensity is a constant over the entire integration, and we assume it as $I_b$.

\par For a fixed pulse longitude, its average flux corresponds to the mean intensity of all radio signals along the line of sight over the entire integration time. In different sub-integrations, the radio emission intensity of each pulse longitude shows slight fluctuations. This fluctuation may be due to the interstellar scintillation. For the whole $360^{\circ}$ of longitude radiative pulsar, the fluctuation intensity in the strong emission regions (e.g., main pulse) becomes the main influence. The effect of the fluctuation can be eliminated for the pulsar if the integration goes to infinity.
\par To determine the baseline intensity of the entire 110-min observation, the baseline intensity response with time is first taken into account. After eliminating the effect of the baseline intensity response with time, we divide the total individual pulses ($N_{\rm period}$) into two equal individual pulses. The number of the individual pulses of the first sub-integration (from 0 to 55-min) is set as $N_1$. The average radio intensity of this pulsar over the whole pulse phase (i.e., $360^{\circ}$ of longitude) is denoted as $I_{e_1}$ and $I_{e_2}$ for the first and second sub-integrations, respectively. We could have,
\begin{equation}
    I_{e_1} = \frac{1}{N_{1}} \sum_{\rm iperiod = 1}^{N_{1}} I_{\rm iperiod},
    \label{A1}
\end{equation}
and
\begin{equation}
    I_{e_2} = \frac{1}{(N_{\rm period} - N_{1})} \sum_{\mathrm{iperiod} = N_{1} + 1}^{N_{\rm period}} I_{\rm iperiod},
    \label{A2}
\end{equation}
where $I_{\rm iperiod}$ correspond to the $\rm iperiod$-th pulse.
\par After subtracting the baseline intensity $I_b$ of this observation, we have two average pulse profiles of the intrinsic radio emission of this pulsar over the two equal sub-integrations (i.e., $(I_{e_1} - I_b)$ and $(I_{e_2} - I_b)$). The $(I_{e_1} - I_b)$ is not equal to $(I_{e_2} - I_b)$ due to the effect of the fluctuation in emission intensity over different sub-integrations. Considering that the average pulse profile of this pulsar is quite stable over long integration, and taking the effect of the fluctuation into account, we find,
\begin{equation}
    I_{e_1} - I_b = \kappa (I_{e_2} - I_b),
\end{equation}
here $\kappa$ is a parameter that reflects the fluctuations in the radio emission intensity of this pulsar over different sub-integrations.

\bibliography{sample631}{}
\bibliographystyle{aasjournal}



\end{document}